\documentstyle[epsf]{mn2e}

\title[The soft excess in AGN]{Can the soft excess in AGN originate from disc reflection?}

\author[C.~Done \&  S.~Nayakshin] {Chris Done$^1$, 
and Sergei Nayakshin$^2$ \\
$^1$Department of Physics, University of Durham, South Road, Durham DH1 3LE,UK\\
$^2$Department of Physics and Astronomy, University of Leicester, 
University Road, Leicester, LE1 7RH
}

\date{Submitted to MNRAS}
\pagerange{\pageref{firstpage}--\pageref{lastpage}} \pubyear{2003}

\begin{document}

\def\aap{A\&A}
\def\apj{ApJ}
\def\apjl{ApJ}
\def\mnras{MNRAS}

\maketitle

\label{firstpage}

\begin{abstract}

The soft excess seen in many AGN is most probably due to partially
ionized material moving at relativistic speeds close to the black
hole. There are currently two potential geometries for this material,
one where it is out of the line of sight, seen via reflection,
e.g. the accretion disc, the other where it is in the line of sight,
seen in absorption e.g. a wind above the disc. Both models require
apparent fine tuning of the ionization parameter of this material in
order to produce the large jump in opacity at $\sim 0.7$~keV
associated with OVII/VIII, as required to make the soft
excess. However, Chevalier et al (2006) show that these states rather
naturally dominate the absorption spectrum for soft X-ray spectra if
the illuminated material is (at least approximately) in pressure
balance. Here we explore whether hydrostatic pressure equilibrium in a
disc can likewise naturally select the required ionization states in
reflection. We find the opposite. The soft excess 
X-ray excess is much weaker in the hydrostatic models than it is in
the constant density models. Since even the constant density models
cannot fit the largest soft excesses seen without the intrinsic
continuum being hidden from view, this means that reflection from a
hydrostatic disc cannot realistically match the data. Even if the disc
structure is instead more like a constant density atmosphere, the
required fine--tuning of the ionization parameter still remains a
problem for reflection models. 

\end{abstract}


\section{Introduction}

The X-ray spectra of Active Galactic Nuclei commonly show a 'soft
excess', a smoothly rising increase in emission below 1~keV
(e.g. Porquet et al 2004). This is unlikely to represent a true
continuum component as it has a remarkably fixed 'temperature' of
$\sim 0.2$~keV in all high mass accretion rate AGN despite a wide
range in black hole mass from $10^{6-9} M_\odot$ (Czerny et al 2003;
Gierlinski \& Done 2004, hereafter GD04; Crummy et al 2006).  Instead,
a fixed energy for the emergence of this component is easier to
produce in atomic rather than continuum processes. In particular,
there is a strong jump in opacity at $\sim 0.7$~keV from partially
ionized material, where OVII/OVIII (and the Fe M shell UTA) combine to
produce much more absorption above this energy than below.  However,
the observed soft excess is smooth, with no discernible strong
characteristic edge and line features expected from atomic processes.
Thus if this is the origin of the soft excess, the partially ionized
material must have strong velocity gradients so that Doppler shifts
smear these out.

There are two ways in which such partial ionisation could produce a
soft excess, either with optically thick material seen in reflection
or optically thin material seen in absorption. These lead to two
different geometries for the material: a disc out of the line of sight
for reflection, and material above the disc in the line of sight for
absorption.

A {\em reflection} geometry sets limits on the size of the soft X-ray
excess to only a factor 2 above the extrapolation of the higher energy
spectrum (assuming an isotropic source above a smooth disc surface) as
the maximum reflected flux is set by the level of the illuminating
flux. However, some sources (predominantly Narrow Line Seyfert 1's)
are observed to have soft excesses which are much stronger than this,
leading to the requirement for anisotropic illumination and/or a
corrugated disc surface which hides most of the intrinsic flux from
sight (Fabian et al 2002; Ballantyne, Turner \& Blaes 2004; Fabian et
al 2005; Merloni et al 2006). An alternative way to implement this is
via a 'lamppost' source geometry, where the height of the source above
the black hole is rather small. Light bending then preferentially
illuminates the inner disc, giving strong velocity smearing and
enhancing the reflected fraction (Fabian et al 2004; Miniutti \&
Fabian 2004; Crummy et al 2006).

The alternative {\em absorption} geometry can equally well produce an
apparent soft excess which can fit the XMM--Newton spectra (GD04;
Chevallier et al 2006; Sobolewska \& Done 2007; Schurch \& Done 2006)
and variability (Gierlinski \& Done 2006; Ponti et al 2006). However,
here there is no clear limit on the size of the soft excess, so no
special geometry is required to produce a large fraction of soft flux
above the extrapolation of the higher energy spectrum (Sobolewska \&
Done 2007).  

Both atomic models for the origin of the soft X-ray excess require the
same basic ionization conditions i.e. partially ionised Oxygen to
produce the big jump in opacity at 0.7~keV (equivalently $\log\xi\sim
3$ where $\xi=L/nr^2$ is the photoionisation parameter) and large
(relativistic) velocity smearing (but see Pounds \& Reeves 2007). The
large velocity smearing is naturally produced in both models if the
material is close to the black hole, but the requirement on the
ionization state appears arbitrary.  However, Chevallier et al (2006)
show that this may arise rather naturally in an  absorption geometry if the
material is in some sort of pressure balance. This is linked to the 
ionization instability which results from X-ray illumination, where
the material makes a very rapid switch between almost completely
ionized to almost completely neutral (Krolik, McKee \& Tarter 1981,
hereafter KMT81).  This transition may disrupt the absorbing cloud
beyond this point, so that the neutral material is strongly clumped. A
line of sight though the cloud includes only the highly ionized front
edge (invisible) and the partially ionized transition region, which
has an average value of $\log\xi= 2.5-3$ across a region with column
of $\sim 10^{22-23}$ cm$^{-2}$ (Chevallier et al 2006).

The true instability is only present for hard illuminating spectra (KMT81).
However, the rapid transition through the partially ionized zone is generic,
so even soft illuminating spectra such as those that are more typical of
NLS1's (Brandt et al 1997) produce the required absorption properties
(Chevallier et al 2006).

Here we investigate this fine-tuning in the reflection model, to see
if the constraints of hydrostatic pressure balance in the disc can
likewise lead to a rather natural selection of this ionisation state.
We find instead the converse. The rapid transition due to the
ionization instability generically limits the partially ionised zone
to a small range in optical depth ($N_H\sim 10^{22-23}$ cm$^{-2}$
corresponds to $\Delta\tau<0.1$) so it cannot dominate over the
entire reflecting photosphere of $\tau\sim$~0--1 (Nayakshin, Kazanas
\& Kallman 2000 hereafter NKK00, Nayakshin \& Kallman 2001 hereafter
NK01; Ballantyne, Ross \& Fabian 2001; Rozanska et al 2002).  Either
the disc is very far from hydrostatic equilibrium, or the soft excess
is not from reflection from the disc.

\vskip -24pt
\section{Reflection from disks in hydrostatic equilibrium}

The vertical density structure of an X-ray illuminated disc in
hydrostatic equilibrium is rather complex. External X-ray illumination
heats the top layer of the to the Compton temperature (Compton heating
balanced by Compton cooling). This temperature is higher then the
initial disc photosphere so it expands, becoming less dense and so the
photoionisation parameter, $\xi=L/nr^2$, is large, and the material is
highly ionized.  Further down in the disc the pressure necessarily
increases if the disc is in hydrostatic equilibrium. The illumination
is slightly weaker due to scattering in the upper layer so the
pressure increase can only be produced by increasing the density,
which increases the importance of bremsstrahlung cooling ($\propto
n^2$ whereas the Compton heating and cooling $\propto n$).  Stronger
cooling and weaker heating means that the equilibrium temperature
falls with depth, so the density increases, and the photoionisation
parameter drops. Eventually this means that not all species are
completely stripped and bound electrons give rise to a dramatic
increase in cooling due to recombination lines, so to a
correspondingly rapid increase in density decrease in photoionisation
parameter. This thermal runaway only stops when the ions are mostly
neutral, stopping the dramatic increase in new cooling transitions
available, since the ion states cannot decrease below the ground state
(NKK00).

The sudden increase in cooling efficiency at the onset of line
transitions happens exactly at the range of ion states which are
required for the soft X-ray excess i.e. OVIII/OVII. To produce
the soft excess from reflection requires that these ion states
predominate over the photosphere of the disc, i.e. that these extend
over $\tau=$~0--1. Yet the rapid transition means that these states
only occupy a small range in $\tau$. In the limit of very hard
illuminating spectra, the transition is a true instability, with
widely differing ion states co--existing at the same pressure (KMT81),
though thermal conduction smooths this out into a rapid transition in
a disc ($\Delta\tau\sim 10^{-3}$ see Appendix A of NKK00). However,
for softer X-ray illuminating spectra this is no longer a true
instability. The transition is still rapid, but these partially
ionized states may extend over a somewhat larger range of
$\Delta\tau$.  Given that the soft excess appear strongest in the
Narrow Line Seyfert 1 Galaxies (NLS1), which generally have steep
X-ray spectra and strong intrinsic disc emission (e.g. Czerny et
al. 2003), it is clearly important to investigate the nature of the
instability under these conditions.

\vskip -24pt
\section{Hydrostatic balance models with steep X-ray illumination}
\label{sec:hydro}

\begin{figure*}
\begin{center}
\begin{tabular}{ccc}
\leavevmode
\epsfxsize=0.3\textwidth \epsfbox{fxfd1_new.ps} &
\epsfxsize=0.3\textwidth \epsfbox{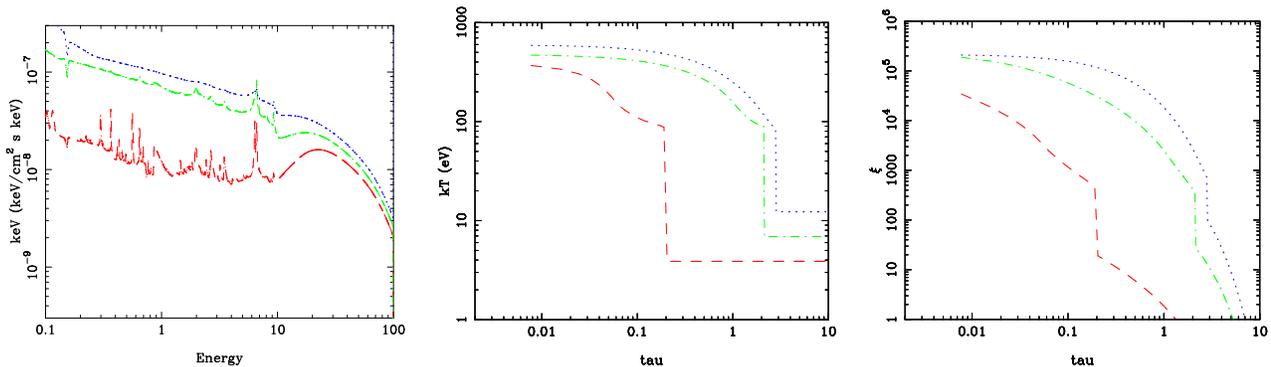} &
\epsfxsize=0.3\textwidth \epsfbox{tau_xi_t.ps}
\end{tabular}
\end{center}
\caption{a) $\nu f_\nu$  spectra for (bottom to top) 
$\dot{m}$ =0.003 (red dashed), 0.03
(green, dot dashed) and 0.3 (blue, dotted) with $F_X/F_{disc}=1$. 
None of these have strong soft excesses.
b) shows the vertical temperature structure for each model. 
The rapid drop in temperature 
marks the onset of line cooling and a rapid rise in density. 
c) shows the corresponding photoionisation structure. Very little of
the disc photosphere can sit at $\xi\sim 10^3$ as this marks the
beginning of the rapid transition.}
\label{fig:fxfd1}
\end{figure*}

\begin{figure*}
\begin{center}
\begin{tabular}{ccc}
\leavevmode
\epsfxsize=0.3\textwidth \epsfbox{fxfd01.ps} &
\epsfxsize=0.3\textwidth \epsfbox{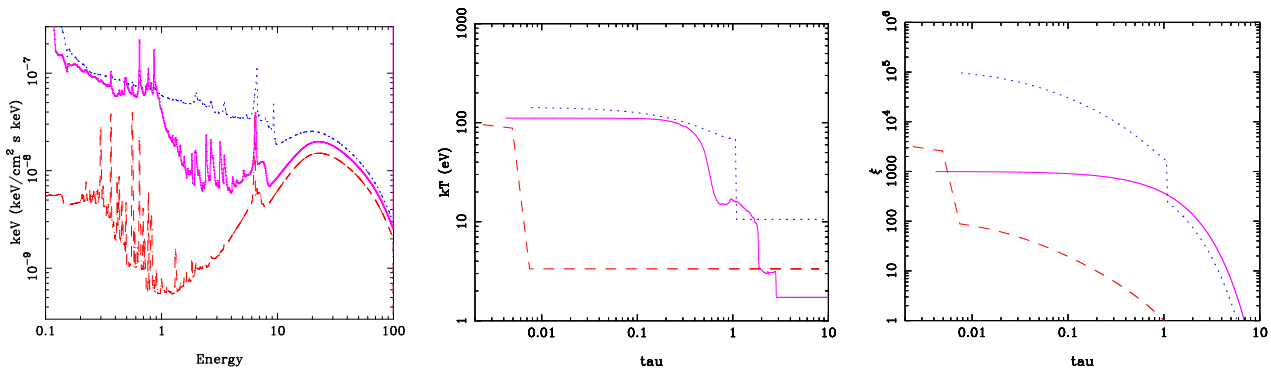} &
\epsfxsize=0.3\textwidth \epsfbox{tau_xi_t_fdfx10.ps}
\end{tabular}
\end{center}
\caption{As for Fig. 1 but with $F_X/F_{disc}=0.1$ for 
$\dot{m}$ =0.003 (red) and 0.3 (blue) compared to a
constant density model with $\xi=10^3$ (magenta, solid). The soft excess is
much more marked in the constant density spectrum as this has
$\xi=10^3$ over the majority of the photosphere of the disc.
}
\label{fig:fxfd1}
\end{figure*}

We use the code of NKK00 to calculate the self-consistent structure of
an X-ray illuminated, hydrostatic disc described by the standard
Shakura \& Sunyaev (1973) $\alpha$ prescription for the viscous
stresses $t_{r\phi}=\alpha P_{tot}$ where $P_{tot}=P_{gas}+P_{rad}$ is
the total (gas plus radiation) pressure. This code makes a number of
simplifying assumptions, such as setting the effective temperature as
a lower limit and calculating line transfer via an escape probability
(see Rozanska et al 2002; Dumont et al 2003). However, it gives
similar results in general (though not in detail) to the two other
codes which can handle this problem using different assumptions
(see Pequignot et al 2001, Rozanska et al 2002).  We take parameters
corresponding to NLS1 galaxies i.e. a steep illuminating X-ray
spectrum with $\Gamma=2.3-2.7$ (with an exponential rollover at 150
keV) and strong intrinsic disk flux, $F_x/F_{disc} \le 1$.

Fig. 1a shows a sequence of reflected spectra at $10R_s$ calculated
with $F_x/F_{disc} = 1$. The three curves are plotted for the
dimensionless accretion rates of $\dot{m} = L/L_{Edd} = 0.3, 0.03$ and
0.003, from top to bottom. There is simply no soft X-ray excess in the
reflected spectrum at the highest accretion rate (blue), which is the
one which should be most applicable to luminous AGN such as Narrow
Line Seyferts. However, the effective viscosity in radiation pressure
dominated discs is not well understood. The stresses could instead
scale more like $\alpha
\sqrt{P_{gas}P_{rad}}$ (Merloni 2003), making the disc denser for a
given $\dot{m}$. The effect of this is similar to simply reducing
$\dot{m}$ in our standard Shakura--Sunyaev disc, but the two models
with smaller accretion rates have soft excesses that are only
marginally larger than that for $\dot{m}=0.3$. None of these models
produce a strong soft excess in reflection.

Fig. 1b shows the temperature structure as a function of vertical
optical depth for each of these accretion rates, while Fig. 1c shows
the corresponding estimate for the photoionisation parameter from
illumination  $\xi(\tau) = 4\pi F_x
\exp(-\tau)/n$, where the factor $\exp(-\tau)$ approximates the loss
of illuminating flux by scattering in the upper layers (appropriate
as these layers are highly ionized), and the diffuse emission is
ignored. Fig. 1c shows immediately why the
soft excess is so weak.  Firstly, the ionization parameter should be
around $\xi \sim 10^3$ to produce the highly but not completely
ionized species of Oxygen (Sobolewska \& Done 2007). In our models
this parameter space occurs over too narrow a region in the optical
depth, just before the temperature drop due to thermal
instability. Secondly, on the top of that layer, there is a skin of
even more highly ionized material, where Oxygen is completely
stripped. Reflection (and emission) from this layer ``fills in'' the
gap in the 0.7--3~keV energy band. As a result, the soft X-ray
emission does not stand out as a well defined rise in the reflected
spectrum.

\vskip -24pt
\section{Comparison with constant density models}\label{sec:constant}

The diluting effect of the highly ionized, upper layer can be removed
by decreasing the size of this layer. This can be done by reducing the
strength of the illuminating flux (NKK00; NK01). Fig. 2 shows results
for two further hydrostatic balance models with $\dot{m}=0.003$ (red)
and $0.3$ (blue) as in Fig. 1, but now with $F_x/F_{disc}=0.1$. The
corresponding temperature and ionisation structures for these are
shown in Fig. 2b and c. These show that the lower illumination can
indeed remove the highly ionized upper layer for the lowest mass
accretion rate, producing a moderate soft excess (Fig. 2a). However,
Fig. 2 also shows results from the code run {\em without} hydrostatic
balance i.e. in constant density mode (magenta) for $\log
\xi_0=\log(4\pi F_x/n)=3$. In order to compare with the results of
Ross \& Fabian (2005) we calculate all constant density models in the
limit of negligible intrinsic disc flux.  The constant density model
with $\log\xi=3$ is much more efficient at producing the soft excess
than any of the hydrostatic models.

Fig 2b shows that the constant density run has a very similar
temperature profile to that of the hydrostatic disc model with
$\dot{m}=0.3$, with an extended region at the Compton temperature of
100-200~eV. However, theier reflected emission is radically different
(Fig 2a). Fig 2c shows that this is because the constant density model
has $\xi\sim 10^3$ over the entire photosphere, from $\tau=0$ to 1,
while the high mass accretion rate hydrostatic disc only has a small
region with this critical ionization parameter, and this region is
buried beneath a a thick, highly ionized layer which dilutes the small
soft excess produced in the region with $\xi=10^3$.

In summary, the ionization parameter is not an independent variable in
a hydrostatic disc. It cannot be set externally as the disc vertical
structure responds to the illuminating flux. The partially ionized
zone with $\xi\sim 10^3$ is near the region where the ionisation state
changes rapidly (see Figs. 1c and 2c). Thus this can only ever be
characteristic of a small portion of the disc photosphere, as opposed
to the constant density models. The small soft excess in the constant
density models can be further suppressed by being buried beneath a
more highly ionized skin, so soft excesses produced by hydrostatic
discs are always much smaller than those predicted by constant density
calculations.

\vskip -24pt
\section{Dependence of results on spectral index}\label{sec:index}

Here we explore the sensitivity of the soft excess to the continuum
shape, comparing the previous results for $\Gamma=2.3$ with steeper
photon indices of $\Gamma=2.5$ and
$\Gamma=2.7$. We fix $F_x/F_{disc}=$~1 and calculate results for 
$\dot{m}=$~0.003 and 0.3.

We parameterize the strength of the soft excess following Sobolewska
\& Done (2007) by convolving the reflected emission with the
relativistic smearing expected from an extreme Kerr disc (the {\sc
laor} profile: Laor 1991). This smooths out individual emission
features and makes soft X-ray features appear as a continuum. The
3--8~keV reflected spectrum is then fit to a power law. Extrapolating
this spectrum down to lower energies, we define the soft excess
strength, SX, as the ratio between the smeared reflected emission and the
extrapolated power law at 0.5~keV.

The open symbols in Fig 3 show the soft excess strength for each of
these hydrostatic disc models, both for the total (incident+reflected)
emission expected from an isotropic source (upper panel) and for the
reflected spectrum alone (lower panel).  Soft excesses in these are
always small, not exceeding even a factor of 1.5 at 0.5~keV for
isotropic emission. This is in stark contrast to the constant density
models calculated from our code with $\xi=10^3$ (filled magenta
symbols), which show much larger soft excesses, close to the
theoretical maximum of a factor $\sim 2$ for isotropic illumination.
We note that the reflected emission produced by our constant density
model is very similar to that of the publically available models of
Ross \& Fabian (2005) for $\Gamma \le 2.3$, but their code gives a
progressively stronger soft excess for steeper illuminating spectra
(see Fig 3), which may be due to their neglect of low ionization
species. However, this is probably not important for most
AGN as the spectral fits of Crummy et al (2006) generally have
$\Gamma\le 2.3$.

\begin{figure}
\begin{center}
\leavevmode
\epsfxsize=0.35\textwidth \epsfbox{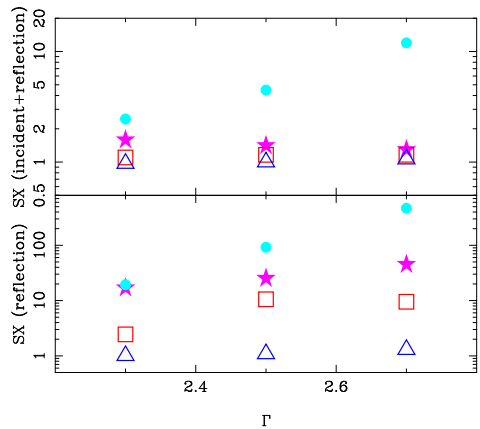}
\end{center}
\caption{The soft excess, SX, defined as the ratio between the
extrapolated power law fit from 3--8~keV to the model at 0.5~keV. The
top panel shows this for isotropic illumination of a flat disc,
while the lower panel shows this for the reflected
emission only. The results from a hydrostatic disc with
$F_x/F_{disc}=1$ are shown as open symbols, with red squares and blue
triangles for $\dot{m}=0.003$ and $0.3$, respectively. The filled
symbols are the constant density models with $\xi=10^3$, with magenta
stars and cyan circles showing results calculated in this paper and in
Ross \& Fabian (2005), respectively. }
\label{fig:sx}
\end{figure}

\vskip -24pt
\section{Discussion}

Both reflection and absorption are expected to be present in the
spectrum at some level, and both of these are easier to identify where
the velocity width is small. There is increasing evidence for
absorption systems with the relativistic velocities required here
(e.g. Pounds \& Page 2006) but these are {\em narrow}.  The
relativistic {\em broadening} (as opposed to blueshift of narrow
features) required in both soft excess models makes it harder to
unambiguously identify characteristic atomic features in reflection or
absorption, and additional physical constraints become important in
assessing their origin.

The strongest reflected soft excesses are from constant density models
with $\xi\sim 10^3$. Even these cannot match the strongest soft
excesses observed with isotropic illumination, hence the requirement
for reflection dominated emission (Fabian et al 2002). The much
smaller soft excesses produced in the hydrostatic disc models mean
that these simply cannot match the largest soft excesses observed in
the data, even with a completely reflection dominated geometry.

Thus the disc cannot be in hydrostatic equilibrium if the soft excess
is made via reflection.  However, it is clear that hydrostatic
equilibrium is not always appropriate. Magnetic flares with
$F_x/F_{disc}\gg 1$ irradiate the local patch of the disc underneath
the flare, causing it to expand {\em sideways} as well as
vertically. Some of the disc material escapes out from underneath the
flare, reducing the optical depth of the skin from that expected in
hydrostatic equilibrium (NKK00, Done \& Nayakshin 2001, Czerny \&
Goosmann 2004).

Such effects are not expected in the more uniform disc illumination
produced in a lamppost model. The entire disc is illuminated, so there
is no unilluminated sector into which the material can escape.
Nonetheless, the hydrostatic equilibrium condition can still be
inappropriate. Firstly, and most fundamentally, the disc may have a
substantial magnetic support, changing the whole behaviour of the
illuminated disc (Blaes et al 2006). Secondly, even without magnetic
fields, hydrostatic balance is disrupted if the disc produces a wind
from the region in which the X--ray reflection spectrum is formed. The
quantitative effect of the wind on the vertical disc structure depends
on the detailed nature of the wind which is not currently well
understood. Nonetheless, mass loss from the disc has the generic
effect of driving the X--ray illuminated disc more towards a constant
density structure.  Thus a powerful wind, such as required by the
absorption model of the soft excess, may also modify the disc vertical
structure so that it can also produce more of a {\em reflected} soft
excess. Nonetheless, the fine--tuning problem still remains, as there
is no obvious physical reason for a magnetically or wind dominated
disc to prefer $\xi=10^3$.

\vskip -24pt
\section{Conclusions}

In this paper we explored the disc reflection origin for the soft
X-ray excess in AGN, using the additional constraint of the
hydrostatic balance on the structure of the illuminated disc
atmosphere. As is well known, material illuminated by X-rays and in
pressure balance is subject to the ionisation instability (KMT81). We
find that, due to this instability, there is a rapid transition
through the partially ionized zone that is required to produce the
soft excess below $\sim 1$ keV. The small optical depth of this zone,
plus the presence of completely ionized material above it mean that
the hydrostatic balance models cannot match the largest soft excesses
seen, even in a reflection dominated geometry where all of the
illuminating continuum is hidden from the line of sight.

Constant density reflection models can have the entire photosphere
dominated by partially ionized material, so produce much stronger soft
excesses. These can match the strongest soft excesses seen in a
reflection dominated geometry, but there is currently no way to
self-consistently account for the fine tuning of the ionization
parameter. 

By contrast, the ionization instability can provide a robust mechanism
to fix the ionization parameter to $\xi\sim 10^3$ in the optically
thin absorption model. The temperature and ionization structure of the
front of the absorber is similar to that in Figs 1b,c and 2b,c so there
is an outer skin of completely ionized material, and then a rapid
transition. The drop in temperature to the more neutral material
implies a rapid density increase at the transition which may fragment
the low ionization part of the cloud into clumps. A line of
sight through the cloud goes through the highly ionized layer (small
opacity), the transition layer (mean $\log\xi\sim 3$) but is not
likely to intercept any of the opaque cool clumps as these have low
covering factor (Chevallier et al 2006). This sets a physical
mechanism for selecting the partially ionized zone, which works even
if the material is not in complete pressure balance. Indeed it seems
likely that any realistic model of the absorption should be 
complex, including adiabatic cooling, spatial inhomogeneity and time
dependent behaviour e.g. if the absorber is formed from cold gas
clumps driven up from the disc being photoevaporated and expanding as
they are exposed to the strong X-ray ionizing flux.

In conclusion, hydrostatic disc models cannot produce the soft excess,
and constant density disk models need fine--tuning as well as
requiring the intrinsic continuum to be suppressed to produce the
largest observed soft excesses. Any real disc 
structure should be between these two limiting cases, hence any
optically thick reflection origin for the soft excess seems unlikely.
Instead, the ionization instability in the absorption model does
provide a robust mechanism for the opacity to be dominated by material
at $\xi\sim 10^3$, and can easily produce even the largest soft
excesses seen without having to hide the intrinsic continuum
Thus it seems most likely that the soft
excess is from absorption, not reflection.

%%% bibliography

\label{lastpage}

\end{document}